\pdfoutput=1

\documentclass[prb,twocolumn]{revtex4-1} 

\usepackage{amsmath}  
\usepackage{amsfonts} 
\usepackage{graphicx} 
\usepackage{caption}
\usepackage{subcaption}
\usepackage{color} 
\usepackage{verbatim}
\newcommand{\di}[2]{\frac{\partial #1}{\partial #2}}

\newcommand{\tdi}[2]{\frac{\textnormal{d} #1}{\textnormal{d} #2}}

\newcommand{\dd}{\textnormal{d}}
\newcommand{\lap}[2]{\mathcal{L} \left\{ {#1} \right\} \left( {#2} \right) }
\newcommand{\lapp}[2]{\mathcal{L} \{ {#1} \} \left( {#2} \right) }
\newcommand{\lappp}[1]{\mathcal{L} \left\{ {#1} \right\} }
\newcommand{\poisson}[2]{\left\{ {#1},{#2} \right\} }

\begin{document}


\title{The classical harmonic chain:\\ solution via Laplace transforms
  and continued fractions}

\author{Nick Kwidzinski}
\author{Ralf Bulla}
\affiliation{Institute for Theoretical Physics, University of Cologne,
  50937 Cologne, Germany}



\date{\today}

\begin{abstract}
The harmonic chain is a classical many-particle system which can be
solved exactly for arbitrary number of particles (at least in
simple cases, such as equal masses and spring constants). A nice
feature of the harmonic chain is that the final result for the
displacements of the individual particles can be easily
understood -- therefore, this example fits well into a
course of classical mechanics for undergraduates.
Here we show how to calculate the displacements by solving
equations of motion for the Laplace transforms $\lap{q_n}{s}$
of the displacements $q_n(t)$. This leads to a continued fraction
representation of the Laplace transforms which can be evaluated
analytically. The inverse Laplace transform of $\lap{q_n}{s}$
finally gives the displacements which generically have the
form of Bessel functions. We also comment on the similarities between this
approach and the Green function method for {\em quantum}
many-particle systems.
\end{abstract}

\maketitle 

\section{Introduction} 
\label{sec:intro}

\begin{figure}[ht!]
\centering
\includegraphics[width=8.0cm]{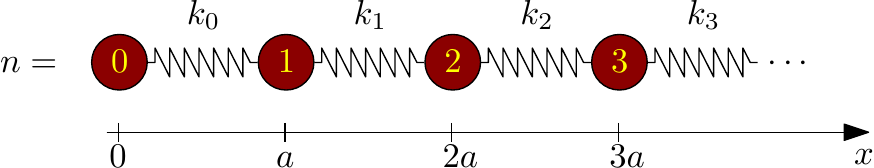}
\caption{Sketch of the harmonic chain, corresponding to the
  Hamiltonian eq.~(\ref{eq:H}).}
\label{harmonic_chain}
\end{figure}

In this paper, we present a method to calculate the time dependence
of the displacements of a classical harmonic chain with the
Hamiltonian 
\begin{equation}
  H = \sum_{n=0}^N \frac{p_n^2}{2m_n} + \sum_{n=0}^{N-1} \frac{1}{2}
  k_n (q_{n+1} - q_n)^2 \ .
\label{eq:H}
\end{equation}
The chain consists of $N+1$ point masses $m_n$ ($n=0,1,\ldots,N$),
connected by springs with spring constants $k_n$ 
(see Fig.~\ref{harmonic_chain}). The displacements $q_n$ in eq.~(\ref{eq:H})
are defined as $q_n = x_n - na$, with $x_n$ the position of mass $n$
and $a$ the lattice constant. The system is in its equilibrium state
 for momenta $p_n =
0$ and displacements $q_n = 0$.

To be specific, let us consider the following initial conditions:
all momenta and displacements are zero except for the
displacement of the oscillator at $n=0$:
\begin{eqnarray*}
   p_n &=& 0 \ , \ n=0,1,\ldots,N \ , \\
   q_0 &\ne& 0   \ , \
   q_n = 0 \ , \ n=1,2,\ldots,N \ .
\end{eqnarray*}
Due to the coupling to the rest of the harmonic chain, the 
left-most oscillator ($n=0$) performs a damped oscillation as
shown in Fig. \ref{semi_q0} below (the result in Fig. \ref{semi_q0} is 
calculated for the semi-infinite chain, i.e.~the limit
$N\to\infty$, with all $m_n=m$, $k_n=k$).

The harmonic chain is an instructive example of a classical
many-particle system which can be solved exactly for any value
of $N$, and it fits nicely into a classical mechanics course
for 1st or 2nd year students. The standard approach is to map
the Newton equations of motion, via a suitable ansatz, to
a problem of linear algebra, that is the calculation of eigenvalues
and eigenvectors of an $(N+1)\times(N+1)$ matrix. This can be done
analytically, at least for simple cases such as equal masses and equal
spring constants.

Here we pursue a different strategy which works as follows:
\begin{itemize}
  \item As a central quantity, we work with the Laplace transform
        $\lap{q_n}{s}$ of the
        time-dependent displacements $q_n(t)$.
  \item For these Laplace transforms, we derive equations of motion
        which involve Poisson brackets $\{\ldots,H \}$ between the
        $q_n$ (or $p_n$) and the Hamiltonian $H$.
  \item Evaluation of these Poisson brackets gives a sequence of
        equations of motion which can be put into the form of a 
        continued fraction.
  \item For equal masses and spring constants, and in the limit $N\to
        \infty$, the continued fraction can be evaluated
        analytically. The spectral function of the  Laplace transform
        $\lap{q_0}{s}$ then acquires a
        semi-elliptic shape, see Fig. \ref{spec}.
  \item Finally, the inverse Laplace transform gives the time-dependent
        function $q_0(t)$ which (for the special case mentioned above) has
        the shape of a Bessel function, see Fig. \ref{semi_q0}.
\end{itemize}
This program appears to be considerably more complex than the standard
approach of mapping to an eigenvalue problem, so what is the benefit
of going through all these steps? We think the main benefit here
is for the students to see various concepts of analytical mechanics --
and theoretical physics in general -- working in an example for which
the physics can be intuitively understood. 
Parts of the derivation such as the calculation
of the Poisson brackets or the derivation of some of the properties
of the Laplace transform might also be used as exercises in problem
classes (while the whole strategy should certainly be explained
in a lecture).

Furthermore, going through the calculation of $q_0(t)$ can serve as a
preparation for more advanced topics of theoretical physics. As we
will discuss in Sec.~\ref{sec:quantum}, there are close 
analogies between
the formalism presented here and the Green function formalism for
quantum many-particle systems. In fact, the calculation of the
Green function for the first site of a tight-binding chain 
very much resembles
the calculation of the Laplace transform
$\lap{q_0}{s}$, and the shape of the
resulting spectral function (the semi-elliptic form in Fig. \ref{spec})
is identical. One of the analogies here is the correspondence between
the Poisson brackets in classical mechanics and the commutators 
in quantum mechanics which results in equations of motion
with a very similar structure.

In Secs.~\ref{sec:laplace}, \ref{sec:eom},
\ref{sec:cont-frac}, and \ref{sec:inverse_laplace}
we follow the steps outlined above with the
final result for $q_0(t)$ discussed in
Sec.~\ref{sec:inverse_laplace}. 
The approach can be generalized to other one-dimensional
geometries, such as the {\em infinite} chain discussed
in Sec.~\ref{sec:generalizations}. For this case, we show
that the displacements $q_n(t)$ are proportional to
the Bessel functions $J_{2n}\left( \sqrt{\frac{4k}{m}}t\right)$, see
eq.~(\ref{solution_inf1}), a result which might also be useful as an
 example of the appearance of Bessel functions
in a physical system.
Section \ref{sec:quantum} is devoted to the analogies to the quantum
case and a summary is given in Sec.~\ref{sec:summary}.

Various issues of the classical harmonic chain have, of
course, been dealt with in the literature, using
approaches different to the one developed here. Goodman
\cite{Goodman}, for example, derives the result for the displacements
$q_n(t)$  -- see eqs.~(\ref{solution_initial_condition_P},\ref{solution_inf2}) employing a normal mode expansion.
The propagation of a localized impulse in a harmonic chain,
a situation which is realized by the initial conditions in
eq.~(\ref{initial_condition_P}), has been discussed in detail by
Merchant and Brill \cite{Merchant}, 
using a similar method. 
An interesting extension of our approach would be to
include external time-dependent forces, which leads to
interesting results as discussed in Cannas and Prato\cite{Cannas}.
They also discussed analogies to the scattering of a quantum particle at a Kronig-Penney potential.

\section{Laplace transform}
\label{sec:laplace}

For a given time-dependent function $f(t)$, we define the
Laplace transform $\lap{f}{s}$ as
\begin{equation}
  \lap{f}{s} = 
  \int_0^\infty {\rm d} t\, e^{-st} f(t) \ , \label{eq:laplace-def}
\end{equation}
with $s=\delta+i\omega$ and $\delta > 0$. The functions $f(t)$ we are dealing with do not grow exponentially (or faster) with time. In this case,
the Laplace transform is an analytic function
in the whole right half of the complex plane. 

As an example, consider the function
\begin{equation}
  f(t) = b\, \cos(\bar{\omega}t) \ ,
  \label{eq:example-bcos}
\end{equation}
which corresponds, in the model introduced in Sec.~\ref{sec:intro}, 
to an undamped oscillation of the displacement $q_0(t)$. The
Laplace transform can be easily evaluated as
\begin{equation}
  \lap{f}{s} = \frac{b}{2} \left[
                 \frac{1}{s-i\bar{\omega}} + \frac{1}{s+i\bar{\omega}}
                 \right] \ .
\end{equation}
It is convenient to illustrate the Laplace transform
via its spectral function defined as
\begin{equation}
S(\omega)=\frac{1}{\pi}\lim\limits_{\delta\rightarrow 0}
\operatorname{Re}\lap{f}{\delta + i\omega} \ .
\end{equation}
With the identity
\begin{equation}
  \lim_{\delta\to 0} \frac{1}{\delta + ix} = 
   -i {\cal P}\left(\frac{1}{x}\right)
                     + \pi \delta(x) \ ,
\end{equation}
(with ${\cal P}$ the principal value and $\delta(x)$ the
$\delta$-function) we obtain
\begin{equation}
  S(\omega) = \frac{b}{2} \left[ \delta(\omega - \bar{\omega}) +
                                  \delta(\omega + \bar{\omega})
                                \right] \ . 
\label{eq:S-discrete}
\end{equation}
In this case, the spectral function is discrete --  a sum of
two $\delta$-functions; performing the inverse Laplace transformation
one can see how these two $\delta$-functions combine again to give
a single oscillator mode, i.e.~the $\cos$-term in 
eq.~(\ref{eq:example-bcos}). For an infinite system, we expect an
infinite number of oscillation modes; the resulting spectral function
then turns out to be continuous, as in the example discussed
in Sec.~\ref{sec:cont-frac}, see eq.~(\ref{eq:spec-semi-ell}).

The properties of the Laplace transform are discussed in detail in
various books, see for example 
Ref.~\onlinecite{AbramowitzStegun}; here is a list of
some of the properties which are used in the following sections:
\begin{itemize}
 \item Initial value and final value theorem: 
 \begin{equation}
 \label{initialvalue1}
 \lim\limits_{s \rightarrow \infty} s \lap{f}{s}=\lim\limits_{t \searrow
   0}f(t) \ .
 \end{equation}
 \begin{equation}
 \label{initialvalue2}
 \lim\limits_{s \searrow 0} s \lap{f}{s}=\lim\limits_{t \rightarrow
   \infty}f(t) \ .
 \end{equation}
Both theorems are valid if the limit exists.
  \item Laplace transform of the integral: 
 \begin{equation}
 \label{integral}
 \frac{1}{s}\lap{f}{s}=\lap{\int_0^t\dd \tau f(\tau)}{s} \ .
 \end{equation}
\end{itemize}
\section{equations of motion}
\label{sec:eom}
In Hamiltonian mechanics, a classical physical system with $N$ degrees
of freedom is described by a set of canonical coordinates
$(q=(q_1,\hdots,q_N),p=(p_1,\hdots,p_N))$. The time evolution is
described by Hamilton's equations
\begin{equation}
	\tdi{p_n}{t}= - \di{H}{q_n},\qquad \tdi{q_n}{t}=\di{H}{p_n} \ .
	\label{Hamiltonian_equations}
\end{equation}
For two functions $f(p,q,t)$ and $g(p,q,t)$ on phase space, 
the Poisson-bracket is defined as 
\begin{equation}
\left\{  f,g\right\} = \sum\limits_{n=1}^N 
\left[ \di{f}{q_n}\di{g}{p_n}-\di{f}{p_n}\di{g}{q_n}\right] \ .
\end{equation}
>From Hamilton's equations (\ref{Hamiltonian_equations}) one obtains 
equations of motion for the total time derivative of a function $f(q,p,t)$
\begin{equation}
\dot{f}=Lf + \di{f}{t} \ ,
\end{equation}
where $L:=\left\{ .,H\right\} $ is the Liouville operator and $\dot{f}$ denotes the total time derivative of $f$.
Consider now 
\begin{equation}
\tdi{}{t}\left[ e^{-st}f(t) \right]=e^{-st}\dot{f}(t)-se^{-st}f(t) \ .
\end{equation}
Performing the integral $\int_0^\infty {\rm d}t \ldots$ on both
sides of this equation gives:
\begin{equation}
\lapp{\dot{f}}{s}=s \lap{f}{s} - f(0) \ .
\end{equation}
This can be used to compute the Laplace transform of higher
derivatives recursively:
\begin{equation}
\lapp{\ddot{f}}{s}=s^2 \lap{f}{s} -s f(0) - \dot{f}(0) \ .
\end{equation}
For explicitly time-independent functions, the two equations of motion
can be rewritten as 
\begin{equation}
	\lap{Lf}{s}=s\lap{f}{s} - f(0) \ ,
	\label{eom1}
\end{equation}
\begin{equation}
	\lap{L^2f}{s} = s^2 \lap{f}{s} -s f(0)- \dot{f}(0) \ .
	\label{eom2}
\end{equation}
The equation of motion eq.~(\ref{eom2}) is the central
equation for the derivation of $\lap{q_n}{s}$, as described in the
following section. In general, the  application of the Liouville
operators in $Lf$ and $L^2f$ generates combinations of the 
coordinates $q_n$ and $p_n$ (depending on the structure of the
Hamiltonian, of course). Repeated application of the equation of
motion eq.~(\ref{eom2}) might therefore lead to a proliferation of the
number of different Laplace transforms $\lap{f}{s}$ and it is a
priori not clear whether the resulting set of equations can be brought
into a closed form. In the semi-infinite chain form of
eq.~(\ref{eq:H}), the set of equations can be closed using
continued fractions, as shown in the following section.


\section{continued fraction for the semi-infinite chain}
\label{sec:cont-frac}
Let us now consider the Hamiltonian eq.~(\ref{eq:H}) in the limit 
$N \rightarrow \infty$ and choose initial conditions 
$q_0(t\!=\!0)=A$, with all other initial displacements and all momenta
set to zero: 
$q_n(0)=A\cdot \delta_{n,0}$, $p_n(0)=0$.

Our aim is to use the equation of motion (\ref{eom2}) to find an
analytical  expression for the  Laplace transform of the displacement
of the zeroth mass point. So the first step is to compute 
$L^2 q_n=\left\{\left\{ q_n, H\right\}, H\right\}$. 
Using the linearity of the Poisson brackets, the relations
\begin{equation}
\poisson{q_n}{p_l}=\delta_{nl},\quad
\poisson{q_n}{q_l}=\poisson{p_n}{p_l}=0 \ ,
\end{equation}
and the product rule $\poisson{f^n}{g}=nf^{n-1}\poisson{f}{g}$,  we
find $Lq_n=\frac{p_n}{m_n}$. We deduce $L^2 q_n=\frac{1}{m_n}
\poisson{p_n}{H}$ which gives  
\begin{equation}
    m_0 L^2 q_0 = k_{0}\left(q_1 - q_{0}\right) \ ,
\label{newtq0}     
\end{equation}
for $n=0$ and
\begin{equation}
	m_n L^2 q_n  	 =   k_{n-1}q_{n-1}+k_{n}q_{n+1} -\left( k_{n-1}+
	k_{n}\right)q_n \ ,
	\label{newtqn}    
\end{equation}
for all $n\neq 0$. Equations (\ref{newtq0}) and (\ref{newtqn}) are the
Newtonian equations of motion and one could in fact have started from
this point. However, we have started from the Hamiltonian to
illustrate the similarities to the quantum case in
Sec.~\ref{sec:quantum}.

Now we define $Q_n(s):= \lap{q_n}{s}$ and plug $L^2 q_n$ into the
equation of motion 
(\ref{eom2}) to get
\begin{equation}
\lap{L^2q_0}{s}=s^2 Q_0(s) - As \ .
\label{rhs}
\end{equation}
The left hand side is readily computed from eq.~(\ref{newtq0}):
\begin{equation}
\lap{L^2q_0}{s}=\frac{k_0}{m_0}\left(Q_1(s)-Q_0(s)\right) \ .
\label{lhs}
\end{equation}
By combining eqs.~(\ref{rhs}) and (\ref{lhs}) we obtain:
\begin{equation}
\left( s^2 + \frac{k_0}{m_0}\right)Q_0(s)-\frac{k_0}{m_0}Q_1(s)=As \ ,
\end{equation}
which can be rearranged to:
\begin{equation}
Q_0(s)= \frac{As}{\frac{k_0}{m_0} +
  s^2-\frac{k_0}{m_0}\frac{Q_1(s)}{Q_0(s)}} \ .
\label{Q_0}
\end{equation}
The Newtonian equation for $n \neq 0$ gives after Laplace transformation
\begin{eqnarray}
&s^2& Q_n(s)=\lap{L^2q_n}{s} =  \nonumber    \\ 
&=&\frac{k_{n-1}}{m_n}Q_{n-1}(s)
+\frac{k_{n}}{m_n}Q_{n+1}(s)   
-\frac{k_{n-1}+k_n}{m_n}Q_n(s) \ .\nonumber    \\ 
\end{eqnarray}
By dividing the equation by $Q_n(s)$ we get:
\begin{equation}
\frac{k_{n-1}}{m_n}\frac{Q_{n-1}(s)}{Q_{n}(s)}=s^2+\frac{k_{n-1}+k_n}{m_n}-
\frac{k_{n}}{m_n}\frac{Q_{n+1}(s)}{Q_{n}(s)} \ .
\end{equation}
Inverting the relation gives us:
\begin{equation}
\frac{Q_{n}(s)}{Q_{n-1}(s)}=\frac{\frac{k_{n-1}}{m_n}}{\frac{k_{n-1}+k_n}{m_n}+s^2-\frac{k_{n}}{m_n}\frac{Q_{n+1}(s)}{Q_n(s)}}
\ . \label{eq:Qn-Qn-1}
\end{equation}
We can now plug this result (for $n=1$) into eq.~(\ref{Q_0}) and iterate, which
leaves us with an expression for $Q_0(s)$ in the form of a continued fraction
\begin{eqnarray}
Q_0(s)= \cfrac{As}{
  \frac{k_0}{m_0}+s^2-\frac{k_0}{m_0}\cfrac{\frac{k_{0}}{m_1}}{\frac{k_0+k_1}{m_1}+s^2-\frac{k_{1}}{m_1}
    \cfrac{\frac{k_{1}}{m_2}}{\frac{k_1+k_2}{m_2}+s^2 - \ddots }}}
\nonumber \\
\label{eq:cf0}
\end{eqnarray}
Note how the structure of the physical system -- the semi-infinite
chain -- translates to the above equation: going along the chain,
starting from site $0$, corresponds to moving to the right within
the continued fraction. 

For the case of equal masses $m_n=m$ and spring constants $k_n=k$, the
continued fraction simplifies to:
\begin{eqnarray}
Q_0(s)= 
\cfrac{As}{
 \frac{k}{m}+s^2-
\cfrac{\frac{k^2}{m^2}}{
\frac{2k}{m}+s^2-\cfrac{\frac{k^2}{m^2}}{\frac{2k}{m} +s^2- \ddots }}
}
=:\frac{As}{x}\ . \nonumber \\ \label{eq:cf1}
\end{eqnarray}
By noticing the periodicity of the continued fraction the auxiliary variable $x$ can be written as
\begin{equation}
x=\frac{k}{m}+s^2-\frac{\frac{k^2}{m^2}}{\frac{k}{m}+x} \ .
\end{equation}
We can now multiply by $\frac{k}{m}+x$ to obtain the quadratic equation
\begin{equation}
x^2-s^2 x-s^2 \frac{k}{m}=0 \ ,
\end{equation}
with the solutions
\begin{equation}
x=\frac{1}{2}s^2\pm\frac{1}{2}s \sqrt{s^2+\frac{4k}{m}} \ .
\label{x}
\end{equation} 
Plugging $x$ back into eq.~(\ref{eq:cf1}) we get
\begin{equation}
Q_0(s)=\frac{2A}{s \pm \sqrt{s^2+\frac{4k}{m}}} \ .
\end{equation}
Making use of the initial value theorem $q_0(0)=\lim\limits_{s\rightarrow\infty}sQ_0(s)$ we get 
\begin{equation}
\lim\limits_{s\rightarrow\infty}sQ_0(s)=\frac{2A}{1\pm 1}\stackrel{!}{=} A
\end{equation}
This tells us that we have to choose the solution with positive sign
\begin{equation}
Q_0(s)=\frac{2A}{s + \sqrt{s^2+\frac{4k}{m}}}=\frac{Am}{2k}\left(\sqrt{s^2+\frac{4k}{m}} -s \right) \ .
\label{L(q_0)}
\end{equation}
The final value theorem tells us that, in the limit $t \to \infty$,
the displacement $q_0(t)$ approaches $0$:
\begin{equation}
\lim\limits_{t\rightarrow\infty} q_0(t)=
\lim\limits_{s\rightarrow 0}sQ_0(s)=0 \ .
\end{equation}

\begin{figure}[ht!]
\centering
\includegraphics[width=7.0cm]{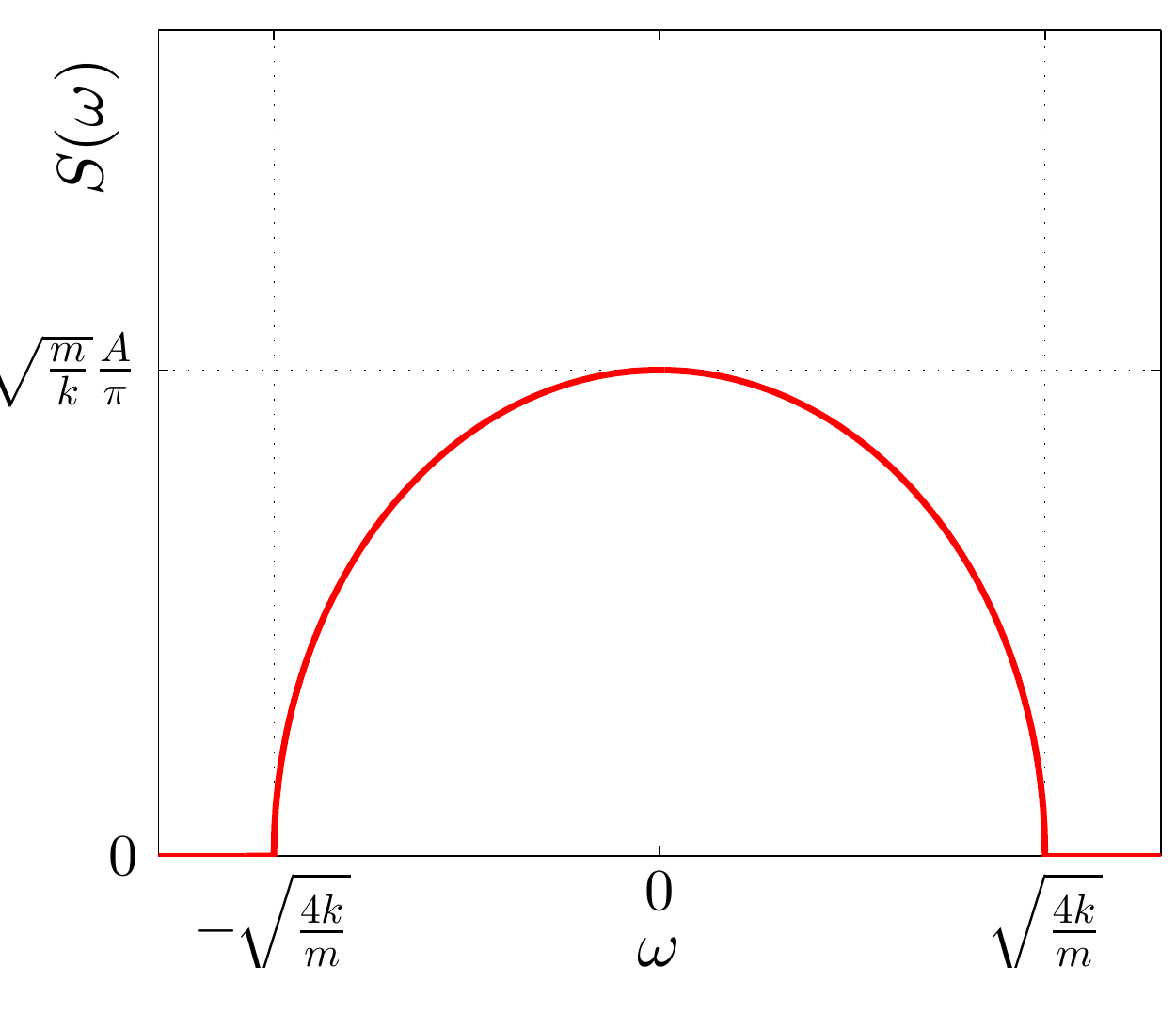}
\caption{The spectral function eq.~(\ref{eq:spec-semi-ell}) of the
  semi-infinite chain assumes a semi-elliptic shape.}
\label{spec}
\end{figure}

From eq.~(\ref{L(q_0)}) we obtain for the spectral function of the semi-infinite
chain:
\begin{equation}
S(\omega)=\frac{Am}{2\pi k}\begin{cases} \sqrt{\frac{4k}{m}-\omega^2} &\mbox{if } |\omega| \leq \sqrt{\frac{4k}{m}} \\
\quad 0 & \mbox{if } |\omega| > \sqrt{\frac{4k}{m}} \ . \end{cases} 
\label{eq:spec-semi-ell}
\end{equation}
The semi-elliptic form of $S(\omega)$ is depicted in Fig.~\ref{spec};
in contrast to eq.~(\ref{eq:S-discrete}) above, we now have a
continuous function in the interval 
$[-\sqrt{\frac{4k}{m}},\sqrt{\frac{4k}{m}}]$.

\section{Inverse Laplace transform}
\label{sec:inverse_laplace}
\begin{figure}[ht!]
\centering
\includegraphics[width=7.0cm]{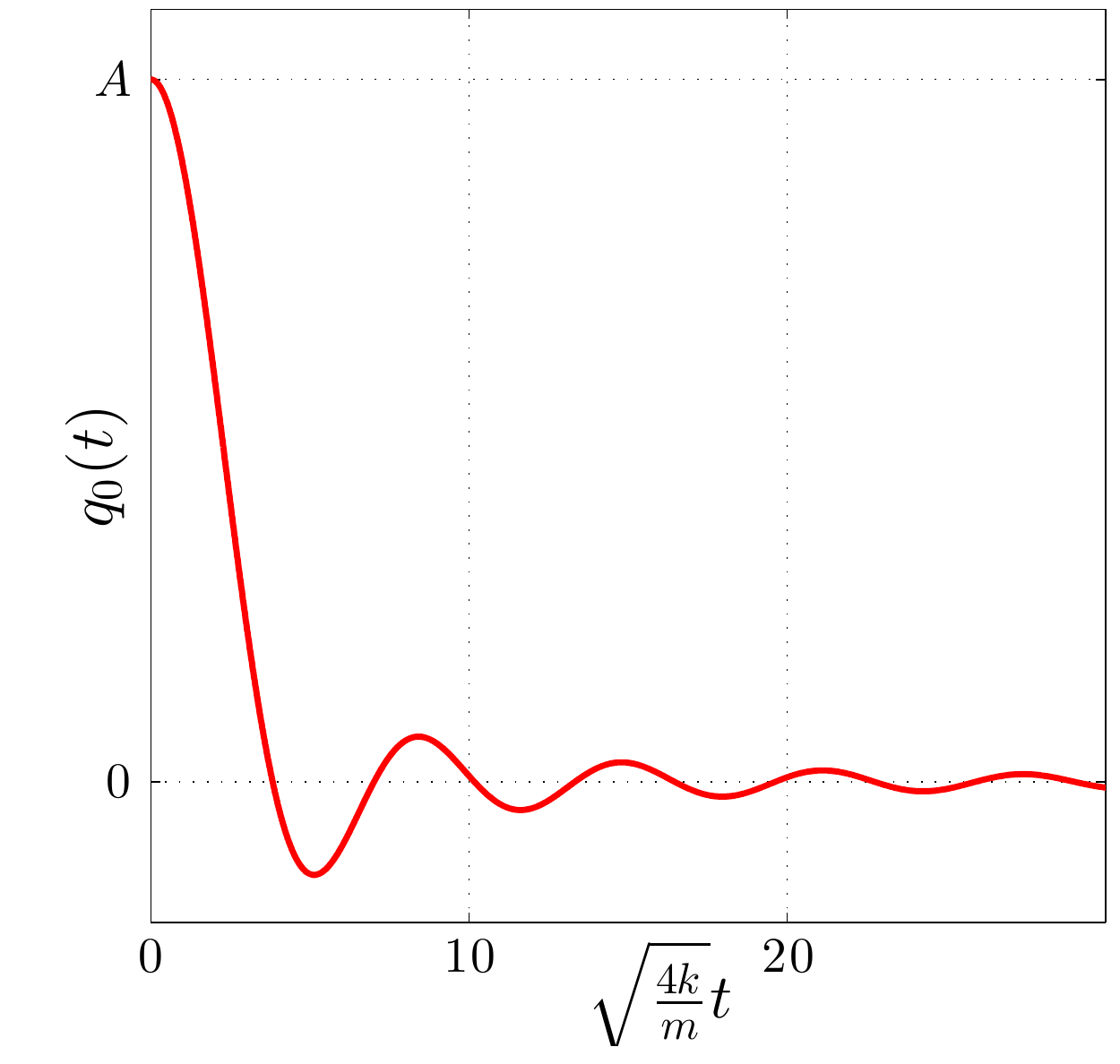}
\caption{Sketch of the solution $q_0(t)$, eq.~(\ref{eq:semi_q0}), for an
  initial value of $q_0(0)=A$.}
\label{semi_q0}
\end{figure}

In order to compute the inverse Laplace transform one can either compute the Bromwich integral or make use of correspondence tables\cite{AbramowitzStegun}. 
A correspondence to (\ref{L(q_0)}) can directly be found in such a table: 
\begin{equation}
	\lappp{q_0}=Q_0=\lappp{\frac{2A}{\sqrt{\frac{4k}{m}}t}J_1\left( \sqrt{\frac{4k}{m}} t \right)} \ ,
\end{equation}
where $J_1$ is the Bessel function of the first kind.
We conclude that the time dependence of the displacement of the zeroth mass point is given by
\begin{equation}
q_0(t)=\frac{2A}{\sqrt{\frac{4k}{m}}t}J_{1}\left(
  \sqrt{\frac{4k}{m}}t\right).
\label{eq:semi_q0}
\end{equation}
As shown in Fig.~\ref{semi_q0}, the displacement $q_0(t)$ describes
the expected damped oscillation of the first oscillator due to the
coupling to the chain.

In what follows the recurrence relations for the Bessel functions\cite{AbramowitzStegun} are required 
\begin{equation}
\label{recurrance1}
	\frac{2n}{\bar{\omega} t}J_n\left(\bar{\omega} t
        \right)=J_{n-1}\left(\bar{\omega} t
        \right)+J_{n+1}\left(\bar{\omega} t \right) \ ,
\end{equation}
\begin{equation}
\label{recurrance2}
	\frac{2}{\bar{\omega}}\di{}{t}J_n\left(\bar{\omega} t
        \right)=J_{n-1}\left(\bar{\omega} t
        \right)-J_{n+1}\left(\bar{\omega} t \right) \ .
\end{equation}
With the help of (\ref{recurrance1}) we can write
\begin{equation}
q_0(t)=AJ_2\left( \sqrt{\frac{4k}{m}}t\right) + AJ_0\left( \sqrt{\frac{4k}{m}}t\right).
\label{solution_initial_condition_A}
\end{equation}
 We then find by solving (\ref{newtq0}) for $q_1$ and using (\ref{recurrance2}) that 
\begin{align}
	q_1(t) & =AJ_4\left( \sqrt{\frac{4k}{m}}t\right) + AJ_2\left(
	 \sqrt{\frac{4k}{m}}t\right)
	 \\
	 & =\frac{6A}{\sqrt{\frac{4k}{m}}t}J_{3}\left(
	 \sqrt{\frac{4k}{m}}t\right).
\end{align}
Using (\ref{newtqn}) as a recursion relation we find for the remaining displacements by induction that
\begin{equation}
q_n(t)=\frac{(4n+2)A}{\sqrt{\frac{4k}{m}}t}J_{2n+1}\left( \sqrt{\frac{4k}{m}}t\right).
\end{equation}

In an analogous calculation one finds for the initial conditions
\begin{equation}
	q_n(0)=0, \quad p_n(0)=P\cdot \delta_{n,0} \quad 
	\text{for } n=0,1,2,\hdots
	\label{initial_condition_P}
\end{equation} 
that
\begin{equation}
Q_0(s)=\frac{1}{s}\frac{2P/m}{s + \sqrt{s^2+\frac{4k}{m}}}.
\end{equation}
Under employment of eq.~(\ref{integral}) one then finds 
\begin{equation}
q_n(t)
=
\frac{(4n+2)P}{m}\int\limits_0^t \dd \tau \frac{J_{2n+1}\left( \sqrt{\frac{4k}{m}}\tau\right)}{\sqrt{\frac{4k}{m}}\tau}.
\label{solution_initial_condition_P}
\end{equation}
In the calculation of (\ref{solution_initial_condition_P}) certain steps where skipped as they are similar to the ones appearing in the calculation of (\ref{solution_initial_condition_A}). 
The final value theorem tells us then that $\lim\limits_{t\rightarrow\infty} q_0(t)=P/\sqrt{km}$. The initial condition (\ref{initial_condition_P}) thus leads to a shift of the equilibrium positions by a factor $P/\sqrt{km}$. The solutions (\ref{solution_initial_condition_A}) and  (\ref{solution_initial_condition_P}) coincide with the solutions one obtains by Goodman's method\cite{Goodman}, the normal mode expansion.

\section{generalizations}
\label{sec:generalizations}
\begin{figure}[ht!]
\includegraphics[width=8.0cm]{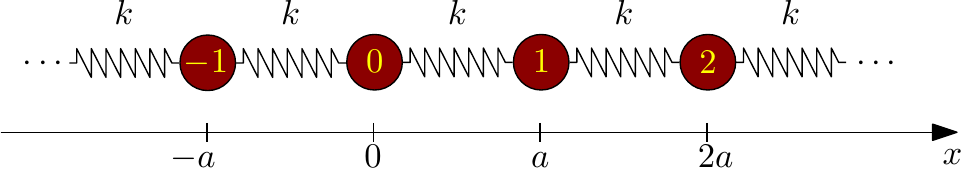}
\caption{Sketch of the infinite harmonic chain, corresponding to the
  Hamiltonian eq.~(\ref{eq:H2}).}
\label{harmonic_chain2}
\end{figure}

In this section, we briefly discuss an infinite chain as sketched in
Fig.~\ref{harmonic_chain2}. The Hamiltonian for this system is given by
\begin{equation}
  H = \sum_{n=-\infty}^\infty \frac{p_n^2}{2m} + \sum_{n=-\infty}^{\infty} \frac{k}{2}
   (q_{n+1} - q_n)^2 \ .
\label{eq:H2}
\end{equation} 
We choose the initial conditions
\begin{equation}
	q_n(0)=A\delta_{n,0}, \quad p_n(0)=0 \quad \text{for all } n \in \mathbb{Z}.
	\label{initial2}
\end{equation}

For simplicity, we take equal masses $m_n=m$ and equal spring constants
$k_n=k$. The symmetry of the system now implies
\begin{equation} 
 q_n(t)=q_{-n}(t) \quad \text{ for all } t \geq 0. 
\label{sym}
\end{equation}  
 To show this explicitly, we define a new set of canonical coordinates by
\begin{align}
	q^{\pm}_n &:= \frac{1}{2}(q_n \pm q_{-n})  & \text{for } n > 0 \\
	p^{\pm}_n &:= p_n \pm p_{-n}  & \text{for } n > 0
\end{align}
%
%
In the new coordinates, $H$ has the form
\begin{align}
H & = \frac{p_0^2}{2m} + \frac{1}{4m}\sum_{n=1}^\infty  \left(p_n^+
  \right)^2 \nonumber \\
  & + \frac{2k}{2}\left[\left( q_0-q_1^+ \right)^2 +  
  \sum_{n=1}^\infty  \left(q_{n+1}^+-q_{n}^+ \right)^2 \right]
\nonumber \\
  & + \frac{1}{4m}\sum_{n=1}^\infty  \left(p_n^-\right)^2 
  +\frac{2k}{2}\left[ \left( q_1^- \right)^2 +
  \sum_{n=1}^\infty  \left(q_{n+1}^--q_{n}^- \right)^2\right]
\end{align}

\begin{figure}[ht!]
\centering
\includegraphics[width=6.4cm]{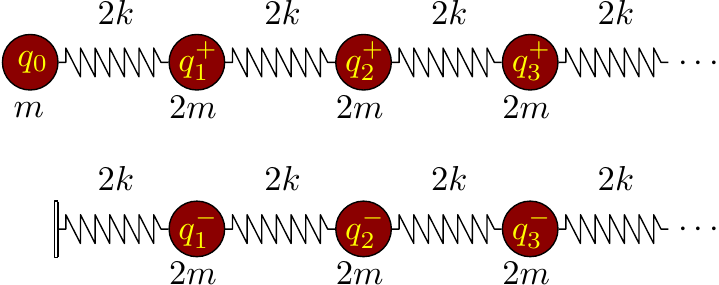}
\caption{Sketch of the harmonic chain, after the canonical
  transformation described in the text. The infinite chain can be
  mapped onto a semi-infinite chain with modified parameters, and a
  decoupled chain.}
\label{H_trafo}
\end{figure}

$H$ now has the form of a Hamiltonian that describes a system of two
decoupled chains as sketched in Fig.~\ref{H_trafo}. The coordinate
$q_0$ couples to a 
semi-infinite chain whose mass points have coordinates $q_n^+$. In
addition, there is a second semi-infinite chain of mass points with
coordinates $q_n^-$ that is decoupled from $q_0$ and obeys fixed
boundary conditions as shown in Fig.~\ref{H_trafo}. It follows
that if we choose initial conditions such that $q_n^-(0)=0$ for all
$n>0$ then $q_n^-(t)=\frac{1}{2}[q_n(t)-q_{-n}(t)]=0$ for all $t \geq
0$. Thus we conclude that the initial conditions (\ref{initial2})
fulfill $q_n(t)=q_{-n}(t)$ at all times.  

\begin{figure}[ht!]
\centering
\includegraphics[width=7.0cm]{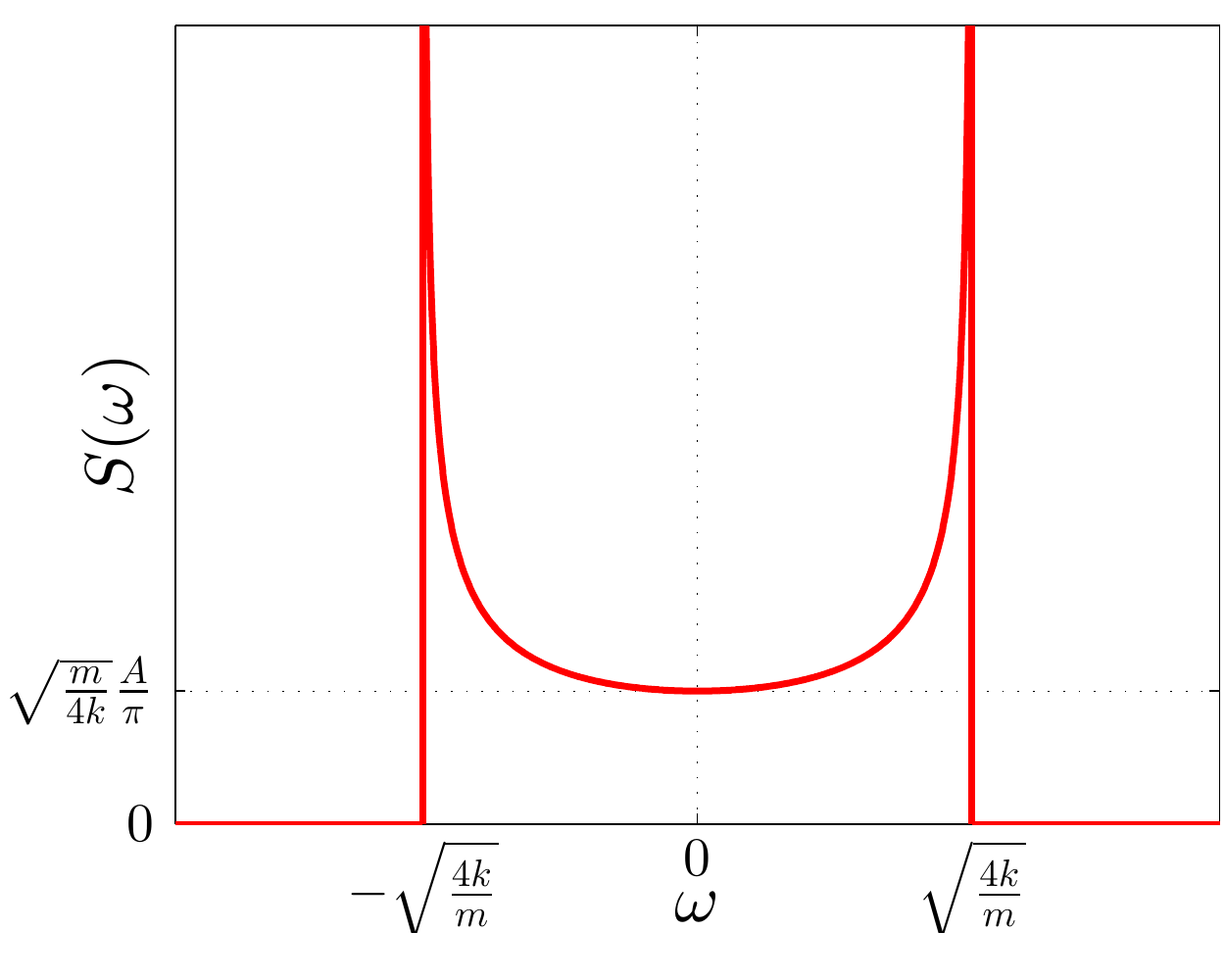}
\caption{The spectral function of the infinite chain, eq.~(\ref{eq:spec-inf})}
\label{spec-inf}
\end{figure}

Moreover, we can simply use the continued fraction eq.~(\ref{eq:cf0})
and plug in the spring constants and masses accordingly, i.e. 
\begin{align}
& k_0=2k,\ k_1=2k,\ k_2=2k,\  \dots \nonumber \\
& m_0=m,\ m_1=2m,\ m_2=2m,\ \dots
\end{align}
to obtain the continued fraction expression for the Laplace transform
of the displacement of the zeroth mass point which then takes the form
\begin{equation}
	Q_0(s) = \cfrac{As}
	{\frac{2k}{m}+s^2 -
	\frac{2k^2}{m^2} 
	\cfrac{1}
	{\frac{2k}{m}+s^2 -
	\frac{k^2}{m^2} 
	\cfrac{1}
	{\frac{2k}{m}+s^2 -
	\frac{k^2}{m^2}\cfrac{1}{\ddots}}}}\ .
\end{equation}
By recognizing the periodicity of the continued fraction, the 
expression for $Q_0(s)$ can be simplified to
\begin{equation}
	Q_0(s) = \pm \frac{A}{\sqrt{\frac{4k}{m}+s^2}}.
\end{equation}
The ambiguity in the sign can again be dissolved by using the initial value theorem, which then provides us with the unique solution
\begin{equation}
	Q_0(s) =  \frac{A}{\sqrt{\frac{4k}{m}+s^2}}.
	\label{LapQ}
\end{equation}
The spectral function for this case reads (see Fig.~\ref{spec-inf}):
\begin{equation}
S(\omega)=\frac{A}{\pi}\begin{cases}
  \frac{1}{\sqrt{\frac{4k}{m}-\omega^2}} &\mbox{if } |\omega| \leq
  \sqrt{\frac{4k}{m}} \ , \\
\quad 0 & \mbox{if } |\omega| > \sqrt{\frac{4k}{m}} \ . \end{cases} 
\label{eq:spec-inf}
\end{equation}

\begin{figure}[ht!]
\centering
\begin{subfigure}[b]{0.5\textwidth} 
\includegraphics[width=7.0cm]{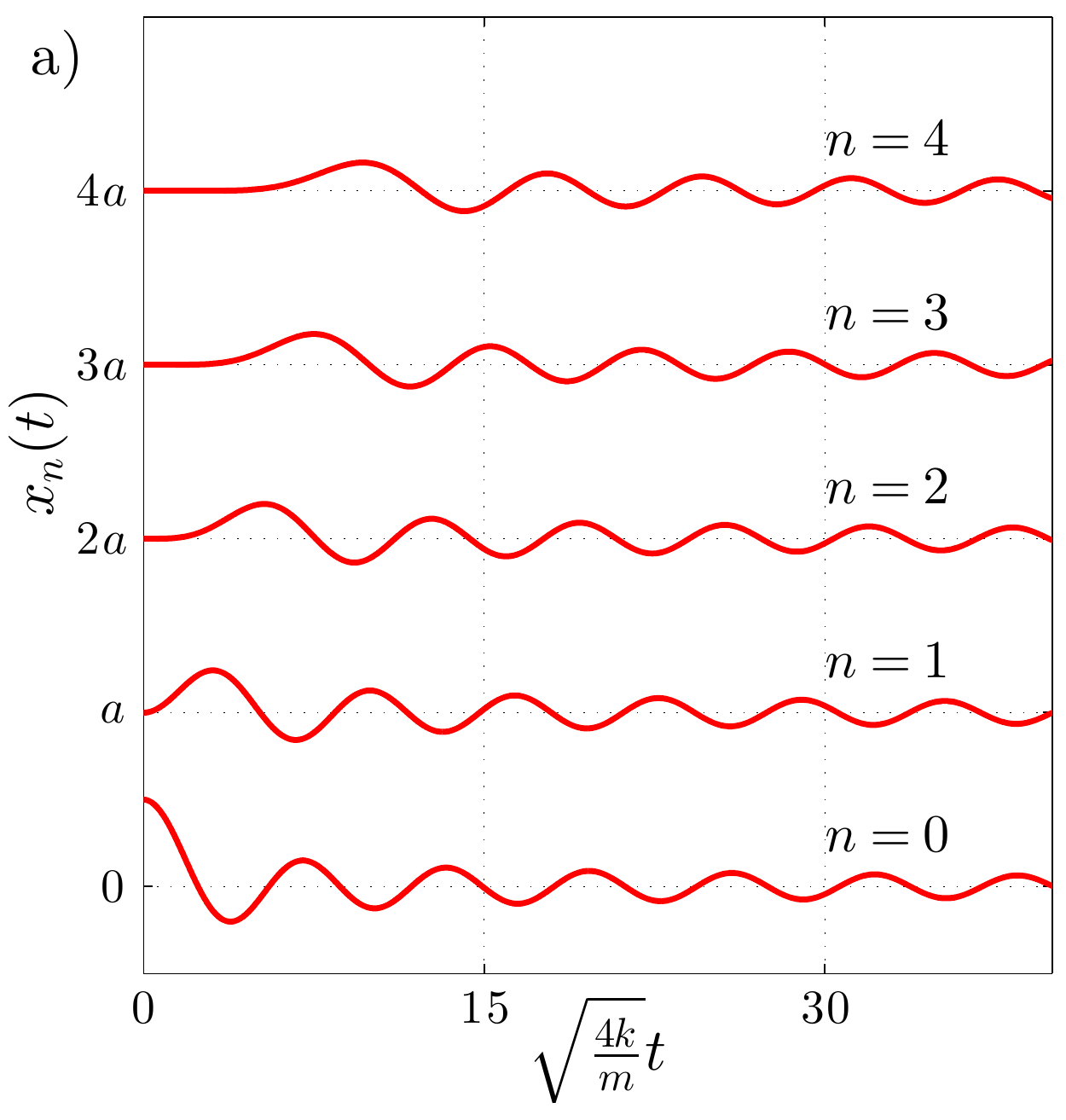} 
\label{fig:gull1} 
\end{subfigure}
\begin{subfigure}[b]{0.5\textwidth} 
\includegraphics[width=7.0cm]{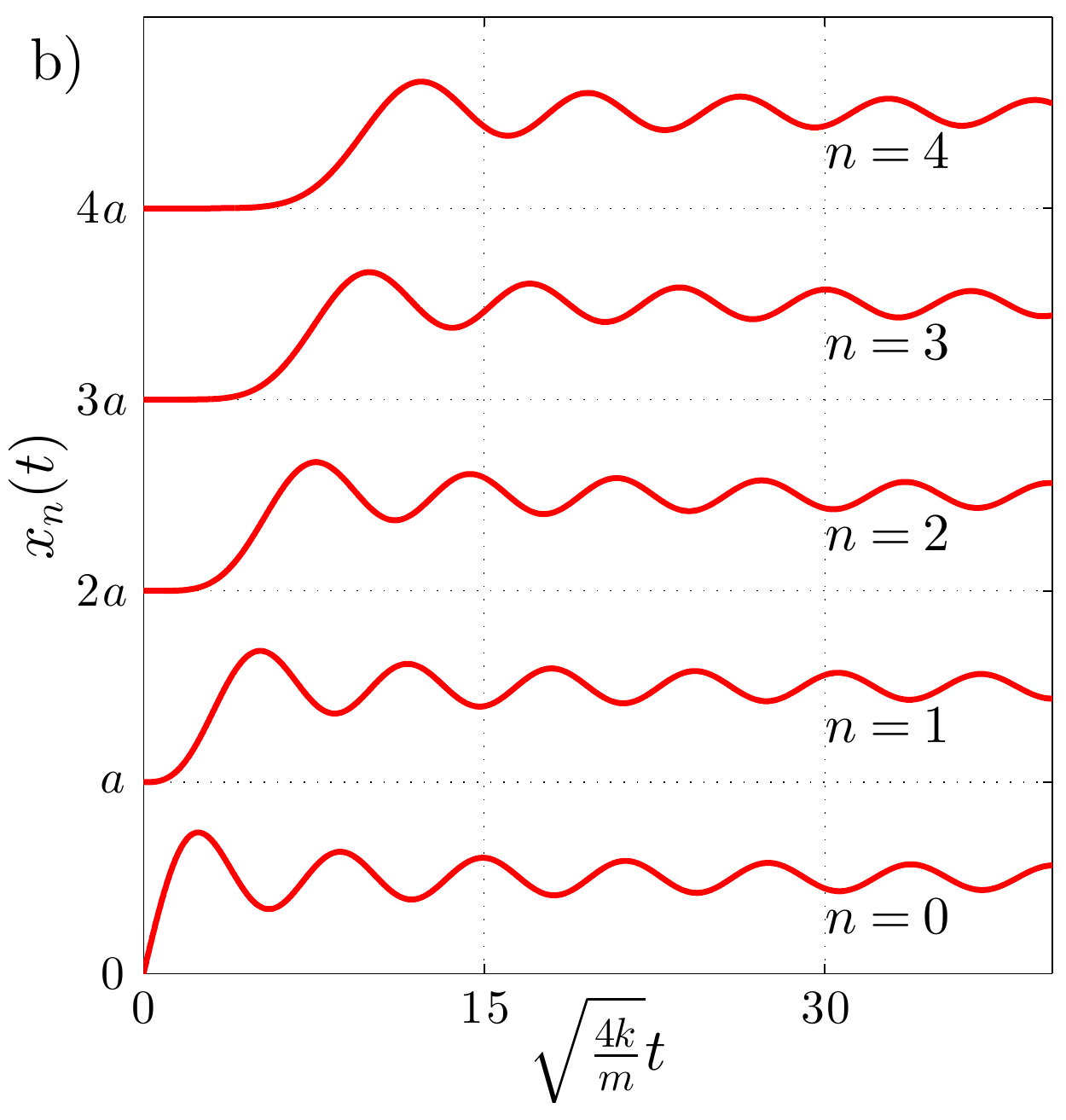} 
\label{fig:gull2} 
\end{subfigure}
\caption{Plot of the positions $x_n(t)=q_n(t)+n a$ as given in
  eqs.~(\ref{solution_inf1}) and (\ref{solution_inf2}) for the initial
  conditions: (a) $q_0(0)=A$ and $p_0(0)=0$, (b) $q_0(0)=0$ and $p_0(0)=P$.}
\label{semi_q0_inf}
\end{figure}

Comparison  of eq.~(\ref{LapQ})  with the Laplace transform of the zeroth Bessel function yields
\begin{equation}
q_0(t)=AJ_{0}\left( \sqrt{\frac{4k}{m}}t\right).
\end{equation}
The Newtonian equations of motion are given by  
\begin{equation}
 \ddot{q}_n		 = \frac{k}{m}\left(q_{n-1} + q_{n+1}\right)-\frac{2k}{m}q_n.
 \label{Newt22} 
\end{equation}
For the case $n=0$ we find by using eq.~(\ref{sym}) that 
\begin{equation}	
 \frac{2k}{m}q_1	 =  \ddot{q}_0	+\frac{2k}{m}q_0.
\end{equation}
Which gives us by employing the recurrence relations (\ref{recurrance1},\ref{recurrance2}) of the Bessel functions that 
\begin{equation}	
 q_1(t)= AJ_2\left( \sqrt{\frac{4k}{m}}t\right).
\end{equation}
Rearranging eq.~(\ref{Newt22}) gives the recursion relation
\begin{equation}
 q_{n+1}		 =\left(\frac{m}{k}\frac{d^2}{dt^2}+2\right)q_n- q_{n-1}.
\end{equation}
Using this and the recurrance relations (\ref{recurrance1},\ref{recurrance2}) we find by induction that
\begin{equation}
q_n(t)=AJ_{2n}\left( \sqrt{\frac{4k}{m}}t\right).
\label{solution_inf1}
\end{equation}
Figure \ref{semi_q0_inf}a shows the positions $x_n(t)=q_n(t)+n a$ for
$n=0,1,\ldots,4$. One can clearly see how the initial displacement of
site $0$ spreads through the chain, with the first maximum for each
$x_n(t)$ moving with the speed of sound of the harmonic chain, while
the individual displacements are damped due to the coupling to the
rest of the chain. Note that this damping (for $n=0$) is significantly
reduced as compared to the semi-infinite case, Fig.~(\ref{semi_q0}).

In an analogous calculation one finds for the initial values 
\begin{equation}
	q_n(0)=0, \quad p_n(0)=P\cdot \delta_{n,0} \quad \text{for all
        } n \in \mathbb{Z}\ ,
\end{equation} 
that the Laplace transform of the zeroth displacement is given by
\begin{equation}
	Q_0(s) =  \frac{P/m}{s\sqrt{\frac{4k}{m}+s^2}}
\end{equation}
from which one finds $\lim\limits_{t\rightarrow\infty} q_0(t)=P/\sqrt{4km}$. 
The displacements are given by
\begin{equation}
q_n(t)
=
\frac{P}{m}\int_0^t \dd \tau J_{2n}\left( \sqrt{\frac{4k}{m}}\tau\right).
\label{solution_inf2}
\end{equation}
The solutions (\ref{solution_inf1}) and  (\ref{solution_inf2}) again
match with those obtained by a normal mode
expansion\cite{Goodman}. Just like in the previous section the choice
of non-vanishing initial momenta leads to a shift of equilibrium
positions (see Fig.~\ref{semi_q0_inf}b),
this time by a factor $P/\sqrt{4km}$ which is half the
factor one obtains in the case of the semi infinite chain.

\section{Comparison to the quantum case}
\label{sec:quantum}

Let us now turn to a {\em quantum} many-particle system which, at
first sight, has not much in common with the {\em classical}
harmonic chain discussed so far. The Hamiltonian of a
one-dimensional tight-binding chain of spinless fermions takes the
form
\begin{equation}
  H= \sum_{n=0}^N \varepsilon_n c_n^\dagger c_n +
     \sum_{n=0}^{N-1} t_n \left( c_n^\dagger c_{n+1} +
                    c_{n+1}^\dagger c_n \right) \ ,
\end{equation}
with $c_n^{(\dagger)}$ fermionic creation (annihilation) operators
and the parameters $\varepsilon_n$ and $t_n$ corresponding to
onsite energies and hopping matrix elements, respectively.

Of particular interest in the theoretical investigation of
quantum many-particle systems are Green functions of the form
\cite{Mahan}
\begin{equation}
  G_{ij}(z) \equiv \langle\!\langle c_i,c_j^\dagger \rangle\!\rangle_z 
   = \int_0^\infty {\rm d} t\,  e^{izt} X_{ij} (t) \ ,
\end{equation}
with 
\begin{equation}
   X_{ij} (t) = -i\theta(t) \langle [c_i(t), c_j^\dagger ]_+\rangle \ .
\end{equation}
The Green function 
$\langle\!\langle c_i,c_j^\dagger \rangle\!\rangle_z$ is given by
the Laplace transform of the time-dependent correlation function
$X_{ij} (t)$ -- note that we use here the usual convention of
quantum many-particle theory (with $z=\omega + i\delta$, $\delta >0$)
different to the Laplace transform as introduced in Sec.~\ref{sec:laplace}.
The Green function is now an analytic function in the whole
{\em upper} complex plane. This convention for the  Laplace transform is related to the previous definition given in  
eq.~(\ref{eq:laplace-def}) via
\begin{equation}
   \int_0^\infty {\rm d} t \ e^{izt} f(t) = -i \lap{f}{s=-iz} \ .
\end{equation}
The equation of motion for the Green function 
$\langle\!\langle A,B \rangle\!\rangle_z$ takes the form
\begin{equation}
    z \langle\!\langle A,B \rangle\!\rangle_z + 
    \langle\!\langle LA,B \rangle\!\rangle_z = 
    \langle [A,B ]_+\rangle  \ ,
\end{equation}
(with $LA := [H,A]_-$) in close analogy to eq.~(\ref{eom1}). In
contrast to the classical case, there is no need for a second order
equation of motion here, simply because the Schr\"odinger equation is
a first order differential equation in time.

Application of the equation of motion to the Green functions
$G_n(z) := \langle\!\langle c_n,c_0^\dagger \rangle\!\rangle_z$
generates a set of equations which can be brought into the form
of a continued fraction, with the final result given by
\begin{equation}
   G_0(z) = \cfrac{1}{z-\varepsilon_0-
            \cfrac{t_0^2}{z-\varepsilon_1-
            \cfrac{t_1^2}{z-\varepsilon_2-\ldots}}} \ .
\end{equation}
For equal $\varepsilon_n=\varepsilon$, $t_n = t$ and in the limit
$N\to\infty$, the continued fraction results in the expression
\begin{equation}
   G_0(z) = \frac{1}{2t^2}\left( z - \varepsilon -
           \sqrt{(z - \varepsilon)^2 - 4t^2} \right) \ .
\end{equation}
The corresponding spectral function is now defined via the
{\em imaginary} part of the Green function:
\begin{equation}
  A(\omega) = - \frac{1}{\pi} \lim_{\delta \to 0} {\rm Im} 
              G(z=\omega + i\delta) \ ,
\end{equation}
which results in
\begin{equation}
  A(\omega) = \frac{1}{2\pi t^2}\begin{cases} 
  \sqrt{4t^2 - (\omega - \varepsilon)^2 }    
            &\mbox{if } |\omega - \varepsilon| \leq 2t \\
\quad 0 & \mbox{if } |\omega- \varepsilon| > 2t \end{cases}
\end{equation}
again giving the semi-elliptic shape as in Fig.~\ref{spec}.

Although we omitted a couple of steps here (such as the calculation
of the commutators $LA$), the analogy to the procedure given in
Sec.~\ref{sec:cont-frac} is obvious.
\\
\section{summary}
\label{sec:summary}
In this paper we presented a method to calculate analytically the
displacements $q_n(t)$ of a classical harmonic chain, with the focus
on semi-infinite and infinite geometries for which various initial
conditions were studied. The calculation proceeds via equations
of motion for the Laplace transforms $Q_n(s) := \lap{q_n}{s}$ and
results in a continued fraction expression for $Q_0(s)$. Finally, the 
displacements $q_n(t)$ are obtained through an inverse Laplace
transformation of the $Q_n(s)$, with the $q_n(t)$ acquiring the form
of Bessel functions.

We believe that the models and the calculations as presented in this paper
would be useful as part of a lecture on classical (analytical) mechanics. While
some of the derivations shown here can certainly be used as exercises
in problem classes, students might also profit from studying the
harmonic chain numerically, i.e.~solving the set of coupled
differential equations for the $q_n(t)$. The numerical results for
either $q_n(t)$ or $Q_0(s)$ (from a subsequent numerical
Laplace transformation) can then be compared with the analytical
expressions.

The concept of using equations of motion for the Laplace transforms 
$\lap{q_n}{s}$ might also serve as a useful preparation for a course
on quantum many-particle systems. As discussed briefly in 
Sec.~\ref{sec:quantum}, there are close analogies between the approach
presented here for a classical harmonic chain and the Green function
approach for a quantum-mechanical tight-binding chain.


\section{acknowledgments}

We would like to thank Matthias Herzkamp for pointing out that the
continued fraction eq.~(\ref{eq:cf1}) can be most efficiently derived by
rearranging the equations of motion as in eqs.~(\ref{Q_0}) and 
(\ref{eq:Qn-Qn-1}). 
Part of this work was funded through the Institutional Strategy of the
University of Cologne within the German Excellence Initiative.


\end{document}